\documentstyle[aps,prl,multicol,floats,epsf]{revtex}
\begin{document}
\draft
\preprint{}
\wideabs{
\title{Effect of Stripes on Electronic States in Underdoped La$_{2-x}$Sr$_x$CuO$_4$}
\author{T. Tohyama, S. Nagai, Y. Shibata, and S. Maekawa}

\address{Institute for Materials Research, Tohoku University,
        Sendai 980-8577, Japan}
\date{Received 4 January 1999}
\maketitle

\begin{abstract}

We investigate the electronic states of underdoped
La$_{2-x}$Sr$_x$CuO$_4$ (LSCO) by using a microscopic model,
i.e., $t$-$t'$-$t''$-$J$ model, containing vertical charge stripes.
The numerically exact diagonalization calculation on small
clusters shows the consistent explanation of the physical
properties in the angle-resolved photoemission, neutron
magnetic scattering and optical conductivity experiments
such as the antiphase domain and quasi-one-dimensional
charge transport.  The pair correlation function
of the $d$-channel is suppressed by the stripes.
These results demonstrate a crucial role of the
stripes in LSCO
\end{abstract}
}
\narrowtext

Since the discovery of high $T_c$ superconductivity,
La$_{2-x}$Sr$_x$CuO$_4$ (LSCO) has been extensively studied
as a typical cuprate superconductor, because it has a simple
crystal structure with single CuO$_2$ layer and the hole density
in the CuO$_2$ plane is changeable in a wide range from $x$=0 to 0.35.
Understanding of the electronic states of LSCO is, therefore, of
crucial importance for the study of the superconductivity.

Recently, Ino {\it et al.}~\cite{Ino2} have performed the
angle-resolved photoemission spectroscopy (ARPES) experiment
on LSCO and have reported that the spectrum near
($\pi$/2,$\pi$/2) along the (0,0)-($\pi$,$\pi$) direction is very
broad and weak in underdoped samples.  The feature is
different from the case of underdoped Bi$_2$Sr$_2$CaCu$_2$O$_{8+\delta}$
(Bi2212) where a sharp peak appears near ($\pi$/2,$\pi$/2)~\cite{Kim}.
The difference between the two typical families may provide
a clue of high $T_c$ superconductivity.  Other experimental data
in LSCO have also revealed anomalous behaviors of
structural~\cite{Saini,Bozin}, electronic~\cite{Ino1,Uchida}
and magnetic properties~\cite{Yamada}.  In particular, at $x$=0.12,
an incomplete phase transition from the low temperature orthorhombic
(LTO) phase to the low temperature tetragonal (LTT) phase
(approximately 10\% in volume) has been observed~\cite{Moodenbaugh}.
At the same hole density, the incommensurate antiferromagnetic (AF)
long-range order has been reported~\cite{Suzuki}.
It is interesting that the LTT phase of Nd-doped LSCO with $x$=0.12,
La$_{1.48}$Nd$_{0.4}$Sr$_{0.12}$CuO$_4$, has shown similar
long-range order accompanied by charge order, which is interpreted
as charge/spin stripe order that consists of vertical charge stripes
and antiphase spin domains~\cite{Tranquada}.  Therefore, such stripes
are expected to play an important role in the family of LSCO.
From the theoretical side, disorder or fluctuation of the stripe
phases has been argued as essential physics of high $T_c$
superconductors~\cite{Emery1,Zaanen,Castellani}.
A phenomenological description of the excitation spectrum has also
been shown~\cite{Salkola}.

In this Letter, the electronic states of LSCO are examined in terms
of the effect of the stripes on various excitation spectra.
We employ a microscopic model with realistic parameters for LSCO
(the $t$-$J$ model with the long-range hoppings) under the presence
of a potential which stabilizes the vertical stripes, simulating
the effect of the LTT fluctuation considered to be favorable for
the stripes~\cite{Saini,Bozin}.  The numerically exact diagonalization
(ED) method is used for small clusters.   We find that the
vertical stripe formation causes the suppression of the
single-particle excitation $A({\bf k},\omega)$, consistent with the
ARPES data in underdoped LSCO~\cite{Ino2}.  This is a direct
demonstration that the vertical stripes actually exist in LSCO.
The effects of the stripes on other quantities (spin correlation,
optical conductivity, and pair correlation) are investigated
and the implication of the results is discussed.

The $t$-$J$ Hamiltonian with long-range hoppings, termed
the $t$-$t'$-$t''$-$J$ model, is
\begin{eqnarray}
H&=& J\sum\limits_{\left<i,j\right>_{1{\rm st}}}
      {{\bf S}_i}\cdot {\bf S}_j
    -t\sum\limits_{\left<i,j\right>_{1{\rm st}} \sigma }
    c_{i\sigma }^\dagger c_{j\sigma } \nonumber \\
&& {} -t'\sum\limits_{\left<i,j\right>_{2{\rm nd}} \sigma }
    c_{i\sigma }^\dagger c_{j\sigma }
     -t''\sum\limits_{\left<i,j\right>_{3{\rm rd}} \sigma }
    c_{i\sigma }^\dagger c_{j\sigma }+{\rm H.c.}\;,
\label{H}
\end{eqnarray}
where the summations $\left< i,j \right>_{1{\rm st}}$,
$\left< i,j \right>_{2{\rm nd}}$ and $\left< i,j \right>_{3{\rm rd}}$
run over first, second and third nearest-neighbor pairs,
respectively.  No double occupancy is allowed, and the rest of
the notation is standard.  Recent analysis of ARPES data has
shown that $t'$ and $t''$ are necessary for understanding
not only the dispersion but also the line shape of the spectral
function~\cite{Kim}.  In LSCO, we estimated the ratio
$t'$/$t$ and $t''$/$t$ to be $-$0.12 and 0.08, respectively,
by fitting the tight-binding (TB) Fermi surface (FS) to the
experimental one in the overdoped sample~\cite{Ino2} on the
assumption that in the overdoped region the FS shape of
the TB band is the same as that of the $t$-$t'$-$t''$-$J$ model.
These numbers are slightly smaller than those for Bi2212,
$t'$/$t$=$-$0.34 and $t''$/$t$=0.23 ($t$=0.35~eV)~\cite{Kim}.
The contribution of apex oxygen to the band dispersion may be
the origin of the difference as discussed by Feiner
{\it et al.}~\cite{Feiner}.
By performing an ED calculation of
$A({\bf k},\omega)$ for a 20-site square lattice in the
$t$-$t'$-$t''$-$J$ model, we have confirmed that the quasiparticle
(QP) at {\bf k}=($\pi$,0) and (0,$\pi$) exists above the Fermi
level in the overdoped region, being consistent with the observed
electronlike FS in the highly overdoped sample~\cite{Ino2}.

In the underdoped region, however, the $t$-$t'$-$t''$-$J$
results show sharp peaks along the (0,0)-($\pi$,$\pi$) direction, being
inconsistent with the ARPES data~\cite{Ino2}.  This implies
the presence of an additional effect, which may be a stripe formation.
It is controversial whether the $t$-$J$ model
(also $t$-$t'$-$t''$-$J$ model) itself has the stripe-type
ground state (GS)~\cite{White,Hellberg,Kobayashi,Tohyama}.
A possible origin of the appearance
of stable stripe phase is due to the presence of the long-range
part of the Coulomb interaction~\cite{Emery2,Seibold} and/or
the coupling to lattice distortions.  In LSCO, the LTT fluctuation
seems to help the latter mechanism~\cite{Saini,Bozin}.
In fact, the LTT structure makes Cu-O bonds along $x$ and
$y$ directions inequivalent, leading to directional distribution of
carriers through the anisotropy of the Madelung potential at in-plane
oxygen sites and that of the hopping amplitude between Cu-O.
To model the directional hole distribution and the tendency toward
the stripe instability as simple as possible, we introduce
a configuration-dependent ``stripe'' potential $V_s$.
The magnitude of $V_s$ is assumed to depend on the number of
holes $n_h$ in each column along the $y$ direction of the lattice.
In the following, we use a $\sqrt{18}\times\sqrt{18}$ cluster
with two holes as an underdoped
system.  The maximum number of holes to be considered here
amounts to three for the final states of
electron removal $A({\bf k},\omega)$.
Noting that the 18-site cluster forms three columns under
the periodic boundary condition (BC), $V_s(n_h)$ in each column
is assumed that $V_s(0)=V_s(1)=0$, $V_s(2)=-2V$,
and $V_s(3)=-3V$, being $V$$>$0.
Examples for two configurations are shown in Fig.~1(a).
$V_s(n_h)$ behaves like an attractive potential
for holes independent of distance between holes.
When the three columns are treated as equivalent,
the translational symmetry is preserved and
thus the size of the Hilbert space can be reduced.
The symmetry with respect to 90$^\circ$ rotation is,
however, broken as expected.
We note that the present form of $V_s(n_h)$ is
available only for the small cluster with up to
three holes~\cite{cluster}.
When we treat systems with more than three holes,
we need to modify the form of $V_s(n_h)$ in order
to distinguish, for example, between the configurations
with a single column filled by four holes and with
two columns filled by two holes each.

\begin{figure}[t]
\epsfxsize=8.0cm
\centerline{\epsffile{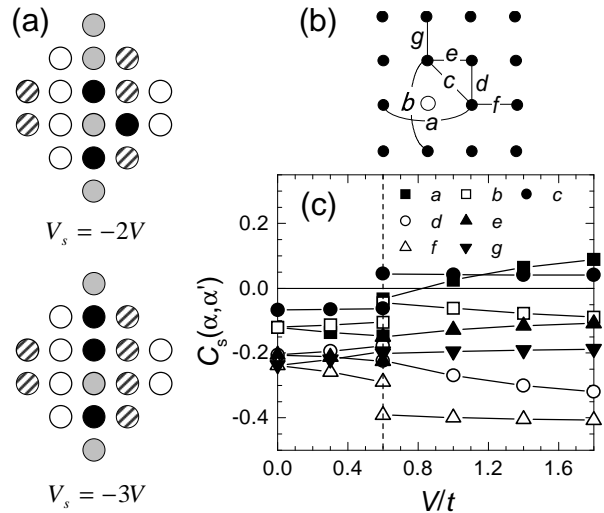}}
\caption{(a) Illustration of the stripe potential $V_s$ for
the 18-site cluster with periodic BC.  The cluster is divided
into three columns represented by the empty, shaded and slash
circles.  When two (three) holes represented by full circles
are on a single column, $V_s=-2V$ ($-3V$).
(b) Labeling configurations used for the calculation of spin
correlation around holes $C_s(\alpha,\alpha')$.
The empty circle denotes the position of a hole.
(c) $C_s(\alpha,\alpha')$ of the $t$-$t'$-$t''$-$J$ model with
the stripe potential $V_s$ on a 18-site 2-hole cluster.
$J$/$t$=0.4, $t'$/$t$=$-$0.12 and $t''$/$t$=0.08.
The discontinuity at $V$/$t$=0.6 is due to a level crossing.}
\label{fig:1}
\end{figure}

Let us start on the GS properties of the 18-site cluster with
two holes as functions of the potential parameter $V$.
The total momentum of the GS is always (0,0), and a level crossing
occurs at $V$/$t$=0.6~\cite{Spin}.  For 0$\le$$V$/$t$$<$0.6
the two holes dominantly form a diagonally bounded pair,
while for $V$/$t$$>$0.6 both holes are mainly on a single column
with two-lattice spacing.  The latter
corresponds to the charge stripes expected in LSCO.

Spin correlation around holes in the GS,
$C_s\left(\alpha,\alpha'\right)\equiv\sum_i <{\rm GS}| n_i^h
{\bf S}_{i+\alpha} \cdot {\bf S}_{i+\alpha'}|{\rm GS}>/N_h$,
is shown in Fig.~1(c).  Here $\alpha$ and $\alpha'$ denote
two sites around a hole following the labeling convention shown
in Fig.~1(b).  $n_i^h$ is the hole-number operator at site $i$,
and $N_h$ is the total number of holes.
In the stripe regime ($V$/$t$$>$0.6), the spin correlation for
the $d$ and $f$ configurations is AF and strong enough to be
comparable to the value of $-$0.33 for the Heisenberg model.
The spin correlation in the $a$ configuration is expected to
be AF due to the hole motion~\cite{Zaanen}.  Our result shows
sizable AF correlation for $V$/$t$$<$0.6.  The magnitude is
reduced when $V$/$t$$>$0.6, since the hole motion
perpendicular to the stripes is suppressed.
However, the correlation is still AF for $V$/$t$$<$0.8.
When $V$/$t$$>$0.8, it becomes weak ferromagnetic (FM).
This FM correlation is, however, due to the periodic BC
imposed on the cluster that makes the same spin orientation
favorable in the $a$ configuration.
In fact, if we use a 5$\times$4 cluster with
open BC along the $x$ direction, the spin correlation across
the stripes is always AF with small magnitude
($\sim$$-$0.05 for $V$/$t$=1.0).  Such AF correlation is
consistent with the experimental observation of the antiphase
spin domains~\cite{Tranquada}.

\begin{figure}[t]
\epsfxsize=8.0cm
\centerline{\epsffile{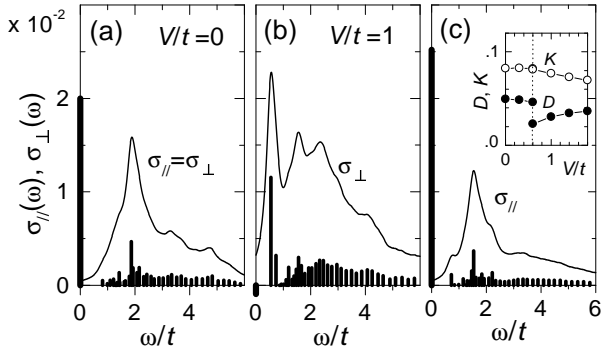}}
\caption{Optical conductivity, $\sigma_{\perp}(\omega)$ and
$\sigma_{||}(\omega)$, of the $t$-$t'$-$t''$-$J$ model
with the stripe potential on a 18-site 2-hole cluster.
$J$/$t$=0.4, $t'$/$t$=$-$0.12 and $t''$/$t$=0.08.
(a) $V$/$t$=0. (b) and (c) $V$/$t$=1.  The delta functions
(vertical bars) for $\omega$/$t$$>$0
are broadened by a Lorentzian with a width
of 0.2$t$ (solid curves).  The Drude weights are shown by
solid bars at $\omega$/$t$=0 multiplied by 0.4.
The negative Drude weight in (b) is due to finite-size
effect of the cluster.  The inset of (c) is the Drude
weight and integrated conductivity averaged over both the
directions as functions of $V$/$t$.}
\label{fig:2}
\end{figure}

Figure~2 shows the optical conductivity $\sigma_{||}(\omega)$
and $\sigma_{\perp}(\omega)$ (parallel and perpendicular
to the stripes, respectively) on the 18-site
cluster with two holes.
When $V$/$t$=1, $\sigma_{||}(\omega)$
shows large Drude weight indicating one-dimensional (1D)
behavior, whereas the Drude part in $\sigma_{\perp}(\omega)$ is
strongly suppressed.  In contrast to ideal 1D
systems~\cite{Stephan}, the sizable incoherent weight remains
in $\sigma_{||}(\omega)$ because the charge carriers in the
stripes are not completely free from the spin configuration
in the spin domains.
The quasi-1D nature along the stripes may explain an
anomalous $x$ dependence of the integrated $\sigma(\omega)$
up to 1.2~eV~\cite{Uchida}: it increases linearly
up to $x$$\sim$0.12 but above $x$$\sim$0.12 the value is
saturated.
At $x$$\sim$0.12, the hole density in each stripe
is about 0.5.  When more holes enter into the stripes,
the kinetic energy of carriers along the stripes
decreases since the energy has a maximum at the density
of 0.5 in the 1D chain with strong correlation.
Thus, the integrated $\sigma_{||}(\omega)$ decreases.
Holes introduced into the spin domains, however, increase
the integrated value.
Therefore, competition between these two effects causes
the saturated behavior above $x$$>$0.12 seen in the experiments.

The inset of Fig.~2(c) shows the total Drude weight $D$ and
the integrated conductivity $K$ averaged over both the
directions.  We find that the difference $K-D$,
which represents the magnitude of the incoherent conductivity,
for $V$/$t$=1 is larger than that for $V$/$t$=0.
The large peak at $\omega$/$t$$\sim$0.6 in Fig.~2(b),
whose energy position is sensitive to $J$, is the origin
of the difference.  These results are consistent
with the fact that midinfrared absorption at around
$\omega\sim$0.27~eV ($\sim$0.7$t$) in LSCO is enhanced
in comparison with the spectra of YBa$_2$CU$_3$O$_{6.6}$ and
Bi2212~\cite{Tajima}.
With further increasing $V$/$t$, the difference $K-D$ decreases.
This behavior may correspond to the suppression of $\sigma(\omega)$
observed in Nd-doped LSCO~\cite{Tajima}.

\begin{figure}[t]
\epsfxsize=8.0cm
\centerline{\epsffile{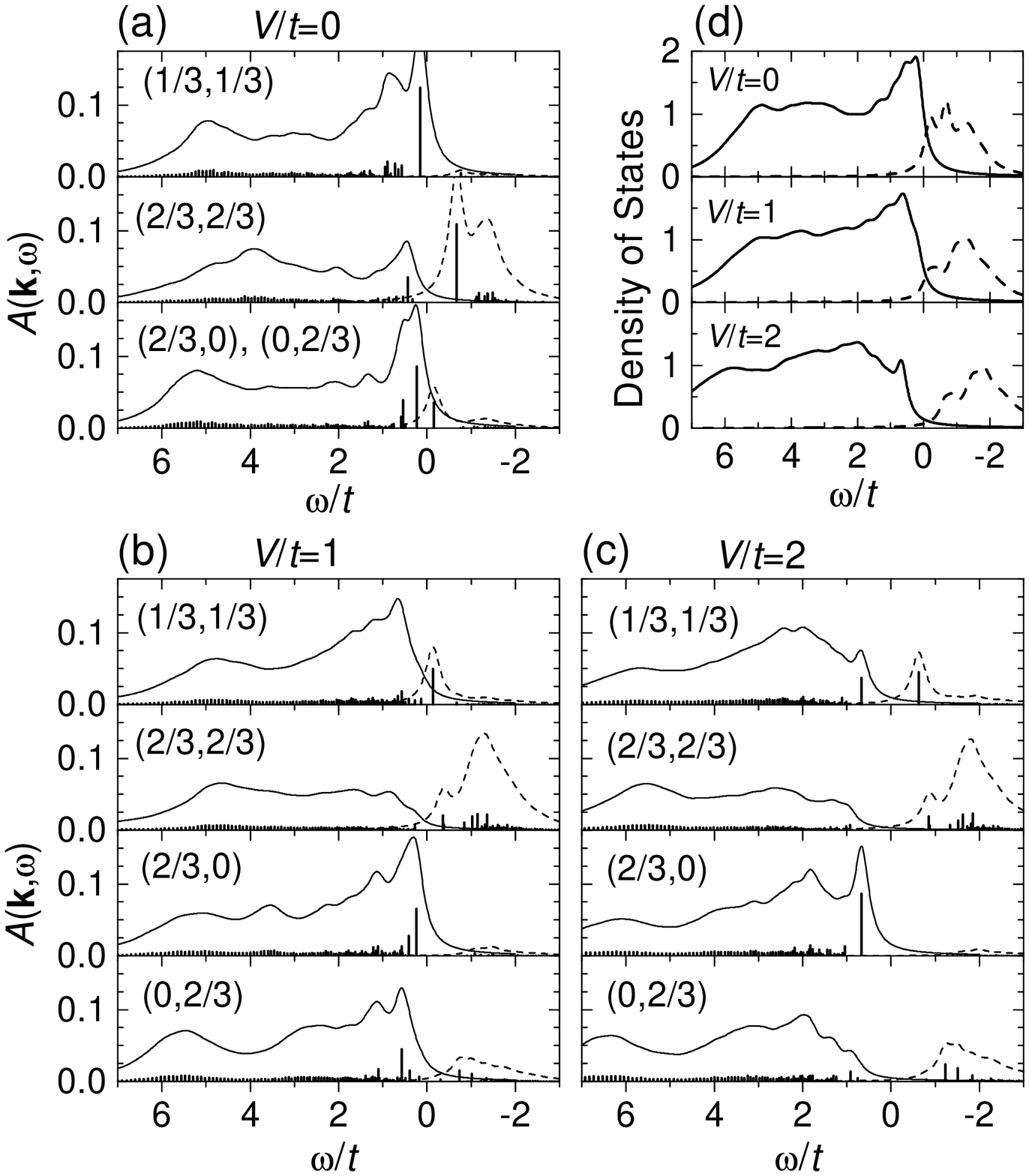}}
\caption{Single-particle spectral function $A({\bf k},\omega)$ of
the $t$-$t'$-$t''$-$J$ model with the stripe potential on a 18-site
2-hole cluster.  $J$/$t$=0.4, $t'$/$t$=$-$0.12 and $t''$/$t$=0.08.
(a), (b), and (c) are results for $V$/$t$=0, 1, and 2, respectively.
The delta functions (vertical bars) are broadened by a Lorentzian
with a width of 0.2$t$ (solid and dashed curves for the
electron removal and addition spectra, respectively).
The momentum is measured in units of $\pi$.
(d) The density of states obtained by $\sum_{\bf k}
A({\bf k},\omega)$.  The zero energy corresponds to the Fermi
level defined as the middle point between the first ionization
and affinity states.}
\label{fig:3}
\end{figure}

Figure~3 shows $A({\bf k},\omega)$ for selected momenta.
When there is no potential [Fig.~3(a)], the large QP peaks
are clearly seen below and above the Fermi level at
($\pi$/3,$\pi$/3) and (2$\pi$/3,2$\pi$/3), respectively.
For $V$/$t$=1 [Fig.~3(b)], however, there is no distinct QP peak
and the spectra become more incoherent.
At ($\pi$/3,$\pi$/3), a peak is seen above the Fermi level.
This may be due to the fact that the stripe potential behaves
like an attractive force between holes.
Here, we note that the reduction of QP weight originating from
the charge stripes is more or less seen at all momenta.
In the presence of the potential,
the coupling between charge carriers in the stripes and neighboring
spin domains becomes weak (see Fig.~2), and the spin
correlation across the stripes is also small [see Fig.~1(c)].
The implication of the two results is an increase of the number of
configurations that participate in the GS.  According to the
configuration interaction picture, such an increase results in
the suppression of the QP weight and the increase of incoherent
structures.  Therefore, the stripe has a
general tendency to make the spectrum broad.
For $V$/$t$=2 [Fig.~3(c)], split-off states from
incoherent structures are clearly seen at $\omega$/$t$=0.7
for ($\pi$/3,$\pi$/3) and (2$\pi$/3,0).  This is due to the
localization of carriers along the direction perpendicular
to the stripes~\cite{Salkola}.  For $V$/$t$=1, the tendency to the
localization may be involved in the low-energy peaks at
(2$\pi$/3,0), but can not be recognized at ($\pi$/3,$\pi$/3).
The broadness of the spectrum at ($\pi$/3,$\pi$/3), i.e.,
along the (0,0)-($\pi$,$\pi$) direction, is thus emphasized
in contrast to the spectra for other momenta.
As a consequence, our results for $V$/$t$=1 explain
the ARPES data~\cite{Ino2,cluster}.
We also found that the broadness cannot be explained by
a diagonal stripe potential.
Figure~3(d) illustrates the density of states (DOS).
For large $V$/$t$,
it shows a gaplike feature near the Fermi level,
which may be related to the suppressed DOS at the
Fermi level observed in the photoemission experiment~\cite{Ino3}.

\begin{figure}[t]
\epsfxsize=8.0cm
\centerline{\epsffile{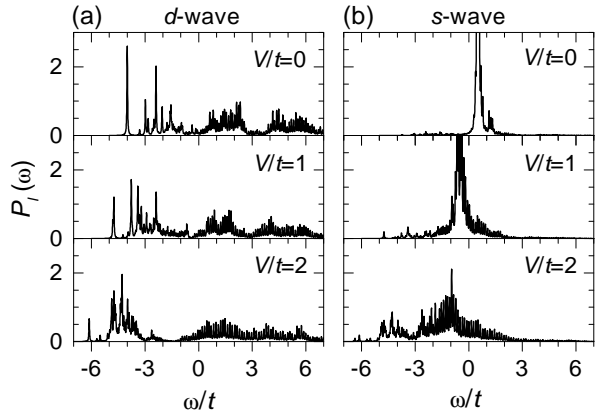}}
\caption{Two-hole pair spectral function $P_l(\omega)$ of
the $t$-$t'$-$t''$-$J$ model with the stripe potential on
a 18-site 2-hole cluster.  $J$/$t$=0.4, $t'$/$t$=$-$0.12 and
$t''$/$t$=0.08. (a) $d_{x^2-y^2}$-wave symmetry ($l$=$d$),
and (b) $s$-wave symmetry ($l$=$s$).  The delta functions
are broadened by a Lorentzian with a width of 0.02$t$.}
\label{fig:4}
\end{figure}

To see the effect of the stripes on the pairing
symmetry of carriers,
we calculate the pair spectral function $P_l(\omega)$ for
two holes added to the Heisenberg ground state~\cite{Poilblanc},
where the hole-pair creation operator is defined by
$N^{-1/2}\sum_{i,\epsilon}\Delta_l(\epsilon)c_{i,\uparrow}
c_{i+\epsilon,\downarrow}$ ($\epsilon$'s are vectors connecting
nearest-neighbor sites) and $\Delta_l(\epsilon)$(=$\pm$1) depends
on $s$- or $d_{x^2-y^2}$-wave symmetry ($l$=$s$ or $d$).
For $V$/$t$=0, the low-energy pair fluctuation is dominated by
$d$-wave symmetry~\cite{Poilblanc} as shown in Fig.~4(a).
With increasing $V$, the low-energy weight of $P_d(\omega)$
decreases, while $P_s(\omega)$ shows enhancement of low-energy
fluctuation.  This implies that the stripes suppress the
$d$-wave pairing accompanied by slight enhancement of
$s$-wave channel~\cite{Emery1}.

In summary, we have investigated the electronic states of
LSCO by using a microscopic model containing the vertical stripes.
The tendency toward the formation of vertical charge stripes
due to the LTT fluctuation was modeled by the stripe
potential, which was added to the $t$-$t'$-$t''$-$J$
model.  The realistic values of $t'$ and $t''$ that
are smaller than those for Bi2212 were used
in the exact diagonalization calculation on small clusters.
Here, we comment that the smallness of $t'$ and $t''$
also works favorably for the stripe formation~\cite{Tohyama}
as compared with Bi2212.
We found that the model explains the ARPES data
with suppressed weight along the (0,0)-($\pi$,$\pi$) direction.
Magnetic properties and charge dynamics are consistent
with experimental data.  The consistent results suggest that
the vertical stripe is an essential ingredient for
the explanation of the physical properties of LSCO.
The pair correlation function of the $d$-wave channel is suppressed
by the stripes, implying the $d$-wave superconductivity is
less favorable under the presence of the stripes.
This may explain why $T_c$ in LSCO is so low, as compared with,
for example, Bi2212.  Our results thus clearly demonstrate
the interesting features of the electronic states caused by the stripes,
and the uniqueness of LSCO among high $T_c$ cuprates.

We thank A. Ino, A. Fujimori, S. Uchida
and Z.-X. Shen for enlightening discussions.
This work was supported by CREST and NEDO.
The numerical calculation were performed in the
supercomputing facilities in ISSP, Univ. of Tokyo, and
IMR, Tohoku Univ.

\medskip



\begin{references}
\bibitem{cluster}In addition to the 18-site cluster, we have
investigated 4$\times$4, 4$\times$5, and 5$\times$4 clusters.
The conclusions in the present study depend on neither the size
of the clusters nor the details of the form of $V_s$,
for example, whether the three columns are equivalent or not.
\bibitem{Spin}At the crossing, the total spin changes from zero
to one.  By examining different clusters in detail, we found
that the change of the total spin is not essential for our results.
\end{references}
\end{document}